\documentclass[conference]{IEEEtran}
\IEEEoverridecommandlockouts

\usepackage{cite}
\usepackage{amsmath, amssymb, mathrsfs, amsfonts,url}
\usepackage{algorithmic}
\usepackage{graphicx}
\usepackage{textcomp}
\usepackage{xcolor}
\usepackage{hyperref}
\usepackage{multirow}
\usepackage{soul}

\usepackage{times,booktabs,tabularx,url}

\definecolor{darkgreen}{rgb}{0.0,0.6,0.0}

\def\BibTeX{{\rm B\kern-.05em{\sc i\kern-.025em b}\kern-.08em
    T\kern-.1667em\lower.7ex\hbox{E}\kern-.125emX}}

\begin{document}

\title{Diff-SAGe: End-to-End Spatial Audio Generation Using Diffusion Models}

\author{
\IEEEauthorblockN{Saksham Singh Kushwaha\textsuperscript{1,*}, Jianbo Ma\textsuperscript{2}, Mark R. P. Thomas\textsuperscript{2}, Yapeng Tian\textsuperscript{1}, Avery Bruni\textsuperscript{2}}
\IEEEauthorblockA{\textsuperscript{1}The University of Texas at Dallas\\
\textsuperscript{2}Dolby Laboratories \\ 
\footnotesize{\textsuperscript{*}Work done during an internship at Dolby Laboratories.}}
}

\maketitle

\begin{abstract}

Spatial audio is a crucial component in creating immersive experiences. Traditional simulation-based approaches to generate spatial audio rely on expertise, have limited scalability, and assume independence between semantic and spatial information. To address these issues, we explore end-to-end spatial audio generation. We introduce and formulate a new task of generating first-order Ambisonics (FOA) given a sound category and sound source spatial location. 
We propose Diff-SAGe, an end-to-end, flow-based diffusion-transformer model for this task. 
Diff-SAGe utilizes a complex spectrogram representation for FOA, preserving the phase information crucial for accurate spatial cues. Additionally, a multi-conditional encoder integrates the input conditions into a unified representation, guiding the generation of FOA waveforms from noise.
Through extensive evaluations on two datasets, we demonstrate that our method consistently outperforms traditional simulation-based baselines across both objective and subjective metrics.

\end{abstract}


\section{Introduction}

\begin{figure}[t]
    \centerline{\includegraphics[width=0.5\textwidth]{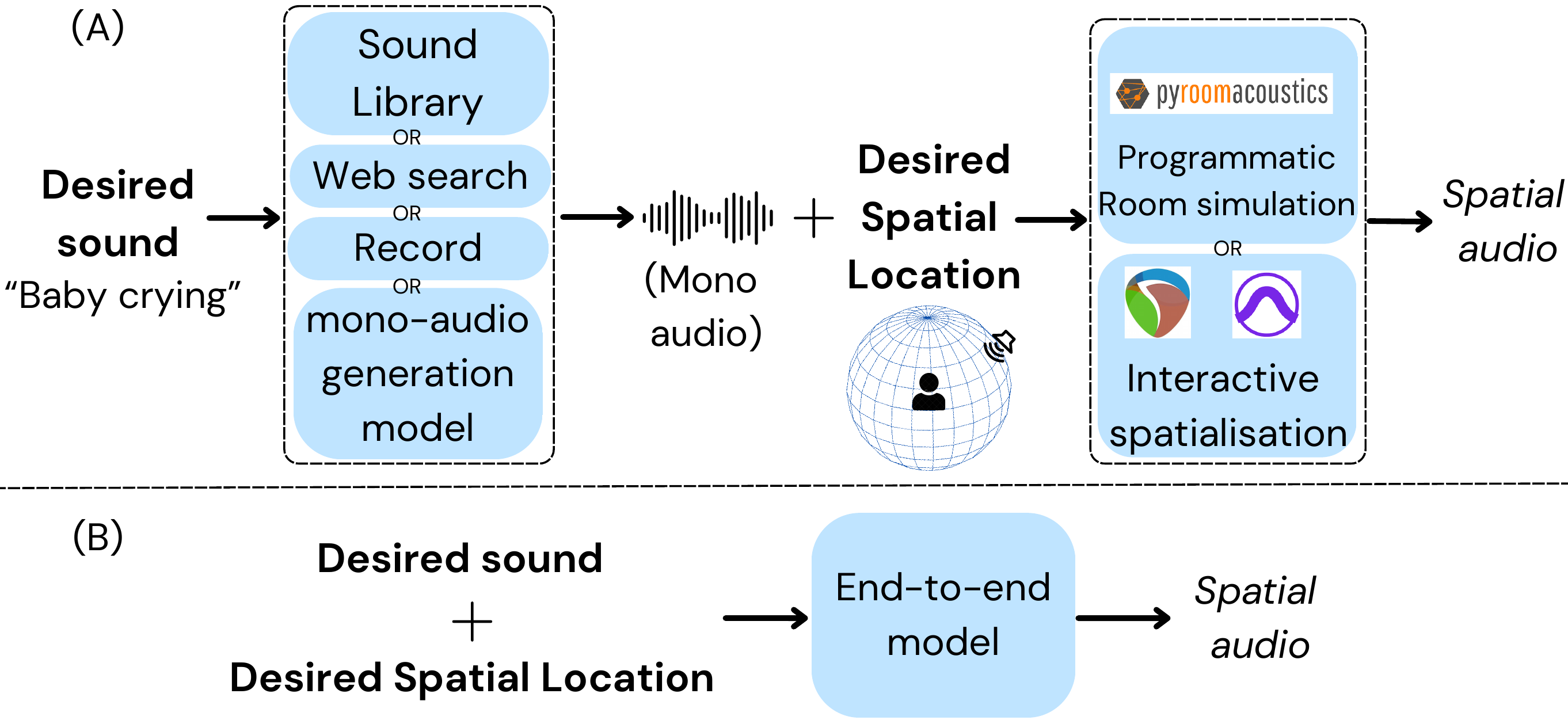}}
    \vspace{-3mm}
    \caption{(A) Traditional simulation-based spatial audio generation. (B) End-to-end spatial audio generation.}
    \label{fig:end2end}
    \vspace{-3mm}
\end{figure}

Spatial audio, including realistic sound and localization cues, is essential for immersive experiences. Its demand is rapidly growing in AR/VR, film, and music, yet authoring high-quality spatial audio remains challenging. Traditional solutions (as shown in Fig.~\ref{fig:end2end}A) that require panning mono audio sources with accompanying spatial metadata are time-consuming~\cite{scheibler2018pyroomacoustics,Tsingos2017}. These methods also assume independence between acoustic content and spatial cues, which is not always true. For example, birdsong tends to be highly directional and emerges from above. Moreover, these methods require expertise to author realistic mixes and struggle to scale for multimodal experiences like visual-to-spatial-audio generation.

End-to-end spatial audio generation offers a promising solution. As shown in Fig.~\ref{fig:end2end}B, it can simultaneously leverage both spatial cues and content information to generate spatial audio directly, bypassing the need for iterative and interactive adjustments. However, this task remains challenging and under-explored. Unlike mono audio, spatial audio involves multiple channels that must represent the semantics while maintaining specific inter-channel relationships corresponding to physically valid source localization.

Previous audio spatialization approaches have focused on augmenting captured mono audio with spatial information from video. For example, \cite{morgado2018self} proposed a self-supervised model for sound source separation and localization to upmix mono audio to Ambisonics synchronized with 360° video. Similarly,~\cite{lin2021exploiting,Gao_2019_CVPR} use visual guidance to upmix mono audio to binaural audio. Unlike previous approaches, we want to natively generate spatial Ambisonics audio, without relying on pre-existing mono audio.

\begin{figure*}[t]
    \centerline{\includegraphics[width=0.9\textwidth]{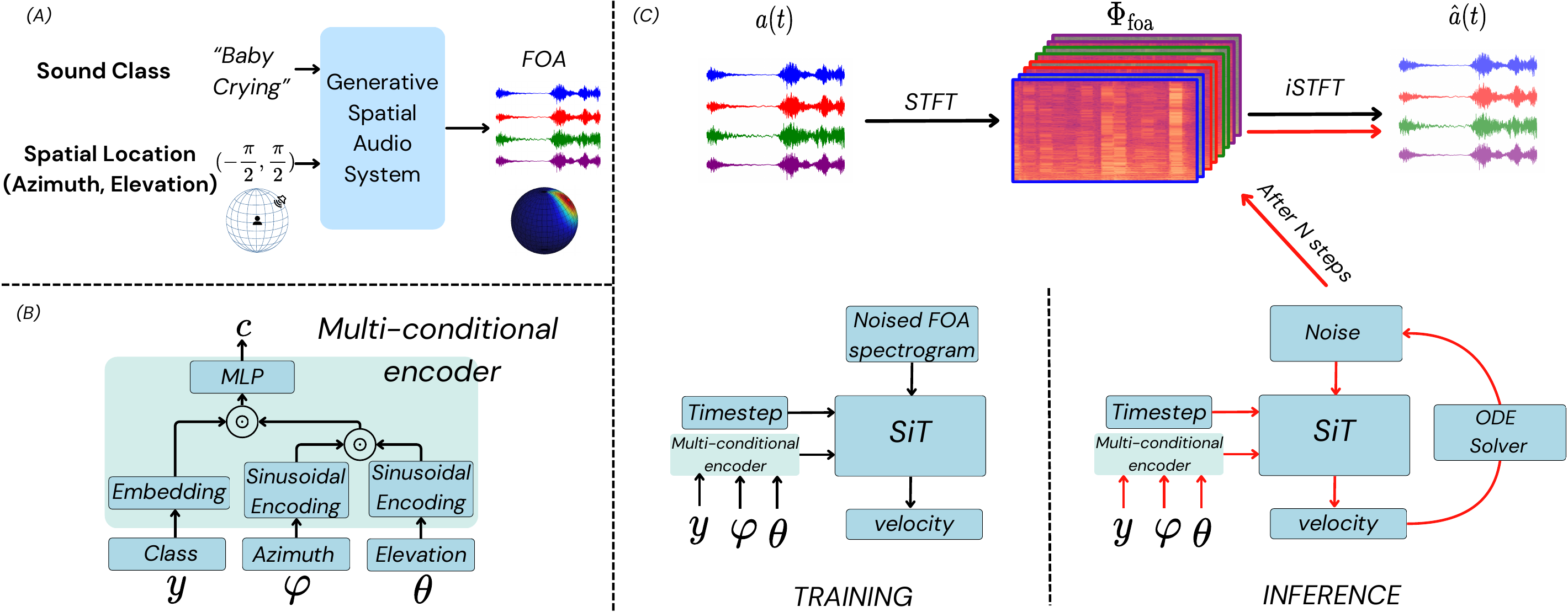}}
    \vspace{-3mm}
    \caption{(A) Our proposed task. (B) Multi-conditional encoder (C) Overall training pipeline of our Diff-SAGe approach. } 
    \vspace{-3mm}
    \label{fig:model}
\end{figure*}

We formulate this problem as generating first-order Ambisonics (FOA) from sound class and sound source location (or Direction-of-arrival). We choose FOA as it is widely accepted due to its flexibility and adaptability~\cite{gerzon1985ambisonics, malham19953,zotter2019Ambisonics}. Following the success of diffusion models for mono audio generation~\cite{ghosal2023text,luo2024diff,liu2023audioldm}, we investigate their potential for our task. Most audio generation models learn to denoise the Mel spectrogram (or its latent) representation of audio, discarding phase information. These models are often conditioned on video, text, or other contextual data inputs. To reconstruct the waveform in the temporal domain, where phase information is required, techniques such as the Griffin-Lim algorithm or vocoders are used to estimate the phase that was not retained in the Mel spectrogram. In spatial audio, phase information is very important, and these existing audio generation techniques fail to reconstruct the inter-channel phase relationships. Hence, a straightforward extension of the mono-audio diffusion model is not feasible.

To approach this task, we propose Diff-SAGe, a flow-based diffusion transformer for generating spatial audio from noise, conditioned on sound class and source location. We overcome the missing phase information in the mel-spectrogram by representing FOA using complex spectrograms. A multi-conditional encoder converts class and location into a unified representation, guiding Diff-SAGe to generate realistic and contextually aligned FOA. We also introduce simulation-based baselines and compare conditional and distributional alignment. Extensive experiments on two datasets demonstrate the superiority of Diff-SAGe over baselines, both in objective metrics and human preference. Our contributions include:
\begin{itemize}
    \item A novel spatial-audio generation task. We formulate this task as generating FOA directly from the class condition and spatial location. To the best of our knowledge, this is the first work in spatial audio generation from scratch.
    \item Diff-SAGe, an end-to-end method for spatial audio generation using a flow-based diffusion transformer. We further introduce simulation baselines and objective and subjective benchmarks for extensive comparison.
    \item Comprehensive testing on two datasets showing the superiority of our proposed approach over baselines.
\end{itemize}

This work establishes a new paradigm in spatial audio generation, paving the way for an exciting field. We include demo material on our companion website. \footnotemark
\footnotetext{\url{https://sakshamsingh1.github.io/spatial_audio_demo.github.io/}}

\section{Method}
In this section, we define the new spatial audio generation task, our proposed Diff-SAGe approach, and the baselines.

\subsection{\textbf{Spatial audio generation task}}

As illustrated in Fig.~\ref{fig:model}A, the objective is to design a generative model that synthesizes FOA audio $a(t)$ based on a given sound category ($y$) and the spatial location of the sound source on a unit sphere. The sound source's position is determined by two parameters: azimuth (horizontal angle) $\varphi \in [-\pi, \pi)$ and elevation (vertical angle) $\theta \in [-\pi/2, \pi/2]$, with the microphone (or listener) located at the origin of the coordinate system. 
This task is complex as the model must capture both the semantic content ($y$) and generate multi-channel audio that is temporally synchronized to accurately encode the spatial location ($\varphi, \theta$) of the source.

\subsection{\textbf{Diff-SAGe: \underline{Diff}usion-based \underline{S}patial \underline{A}udio \underline{Ge}neration}}

Our approach consists of three major components: 1) Multi-conditional encoder, 2) Spatial Audio Diffusion-transformer, and 3) FOA encoder and decoder. We use a flow-based transformer diffusion model (SiT) to generate FOA $\hat{a}(t)$ from the input conditions ($y$,$\varphi$,$\theta$). The multi-conditional encoder creates a unified condition $c$ that is used to generate the FOA complex spectrogram $\mathbf{\hat{\Phi}}_{foa}$ and transformed into an FOA waveform via inverse short-time Fourier transform (ISTFT). 

\subsubsection{\underline{Multi-conditional encoder}} The multi-conditional encoder is used to convert ($y$,$\varphi$,$\theta$) to a single representation $c$ as shown in Fig.~\ref{fig:model}B. More specifically, the sound source belongs to a set of classes, for which we generate a label embedding. $\varphi$ and $\theta$ are transformed into a sinusoidal representation and concatenated. This result is then concatenated with the class embedding and passed through an MLP layer to get a unified condition $c$.

\subsubsection{\underline{FOA representation}}
FOA, a spatial audio format, encodes 3D sound field directionality into four channels:
$\boldsymbol{a}(t) = (a_w(t), a_y(t), a_z(t), a_x(t))$. These represent the omnidirectional  component $a_w(t)$, and the $x$, $y$, and $z$ directional components: $a_x(t)$, $a_y(t)$, and $a_z(t)$, respectively. Most mono audio generation models~\cite{liu2023audioldm,huang2023make,NEURIPS2023_e1b619a9} rely solely on magnitude spectrograms, using a vocoder (such as HiFi-GAN~\cite{NEURIPS2020_c5d73680}) to estimate the phase during waveform reconstruction. This approach cannot be directly extended to FOA as it discards vital inter-channel phase information. Our initial experiments confirm this, which suggests this task is more challenging.

As shown in Fig.~\ref{fig:model}C, our solution is that each FOA channel is represented using complex spectrograms, where the real($R$) and imaginary($I$) parts of the spectrogram are stacked sequentially as $\Phi_i(t, \omega) = [\Phi^R_i(t, \omega); \Phi^I_i(t, \omega)]$, with $i \in \{w, x, y, z\}$. This results in a total of 8 spectrograms. Our FOA representation $\mathbf{\Phi_{foa}}$ is thus defined as:
\[
\mathbf{\Phi_{foa}} = [\Phi_w(t, \omega); \Phi_y(t, \omega); \Phi_z(t, \omega); \Phi_x(t, \omega)],
\]
where $\mathbf{\Phi_{foa}} \in \mathbb{R}^{8 \times T \times F}$, where $T$ is the number of time steps, and $F$ is the number of frequency bins.

\subsubsection{\underline{Spatial Audio SiT}}

Transformer diffusion models~\cite{peebles2023scalable,chen2023pixart} have demonstrated superior scalability and performance compared to the standard U-Net architecture with convolutions. Recent work~\cite{ma2024sit,gao2024lumin-t2x} also highlights the advantages of flow-matching formulations over traditional Denoising Diffusion Probabilistic Models as it offers a simple alternative by linearly interpolating between noise and data. Thus, we employ a flow-based diffusion transformer~\cite{ma2024sit} for this task. Let the data be denoted as $x \sim p(x)$. In our case, $x$ represents $\mathbf{\Phi_{\text{foa}}}$, and Gaussian noise $\epsilon \sim \mathcal{N}(0, I)$, an interpolation-based forward process is defined: \
\begin{equation}
    x_t = \alpha_t x + \beta_t \epsilon,
\end{equation}
where $\alpha_0 = 1$, $\beta_0 = 0$, $\alpha_1 = 0$, and $\beta_1 = 1$ to satisfy this interpolation on $t \in [0,1]$ between $x_0 = x$ and $x_1 = \epsilon$. In our framework, we adopt the linear interpolation schedule between noise and spectrogram data, i.e., $    x_t = tx + (1-t)\epsilon.$

This formulation indicates a uniform transformation with constant velocity between data and noise. The corresponding time-dependent velocity field is given by
\begin{align}
    v_t(x_t) &= \dot\alpha_tx + \dot\beta_t\epsilon \\ 
    &= x - \epsilon, \label{eq:vt_linear}
\end{align}
where $\dot\alpha$ and $\dot\beta$ denote time derivative of $\alpha$ and $\beta$. This time-dependent velocity field $v : [0,1] \times \mathbb{R}^d \to \mathbb{R}^d$ defines an ordinary differential equation named Probability Flow ODE:
\begin{equation}
\label{eq:ode}
    dx = v_t(x_t)dt.
\end{equation}
We use $\psi_t(x)$ to represent the solution of the Probability Flow ODE with the initial condition $\psi_0(x) = x$. By solving this Probability Flow ODE from $t=0$ to $t=1$, we can transform noise into a data sample using the approximated velocity fields $v_{\theta}(x_t, t)$. During training, the flow-matching objective directly regresses the target velocity:
\begin{equation}
    \mathcal{L}_{v} = \int_0^1 \mathbb{E}[\parallel v_{\theta}(x_t, t) - \dot\alpha_tx - \dot\beta_t\epsilon \parallel^2]dt,
\label{eq:loss}
\end{equation}

Given the FOA spectrogram, we flatten it using 2$\times$2 patches and apply a linear interpolation schedule to generate the input and target data. During training, each FOA sample is paired with its corresponding class label and spatial location, which are encoded together. We then apply a regression loss between the predicted and ground truth velocities as in Eq.~\eqref{eq:loss}. 

During training, classifier-free guidance is used and condition $c$ is masked with a null token $\emptyset$ with probability $p$. During sampling, the model computes velocity as $v_{\theta}^{\zeta}(x, t; c) = \zeta v_{\theta}(x, t; c) + (1 - \zeta)v_{\theta}(x, t; \emptyset)$ for a fixed $\zeta > 0$. The training and inference steps are illustrated in Fig.~\ref{fig:model}C.

\subsection{\textbf{Simulation Baselines}}
Traditional approaches generate spatial audio by placing a sound source and listener at the desired spatial locations in a virtual environment and synthesizing corresponding spatial cues. The first step consists of finding mono audio recordings matching the desired semantics. Given class category $y$, we use corresponding real mono recordings or text-to-(mono)audio generation models conditioned on `A sound of \{class-label\}'. Using \texttt{pyroomacoustics}~\cite{scheibler2018pyroomacoustics}, we simulate room impulse responses (RIRs) for a tetrahedral microphone array of cardioid capsules placed in the center a large, shoebox room of dimensions $30m\times20m\times10m$. The sound source is placed at a 1m distance from the microphone array in the desired direction ($\varphi,\theta$). The simulated microphone signals are synthesized by convolving the source signal with the RIRs, then converted to FOA using the \texttt{spaudiopy} library~\cite{hold2019spatial}.

We use three sources of mono audio segments for our simulation baselines: reference audio, AudioLDM~\cite{liu2023audioldm}, and Tango~\cite{ghosal2023text}.
\section{Experimental Setup}

\begin{table*}[t]
\centering
\caption{\normalfont{Comparison between Diff-SAGe and baseline approaches. Evaluation is conducted on the test set of TAU-19 and TAU-NIGENS-20.}}
\vspace{-3mm}
\label{tab:main}
\scalebox{1.0}{
\begin{tabular}{cc|ccccc|ccccc}
\toprule
 & & \multicolumn{4}{c}{TAU-19} & & \multicolumn{4}{c}{TAU-NIGENS-20} & \\
 & & \multicolumn{2}{c}{Condition} & \multicolumn{3}{c|}{Distribution alignment} & \multicolumn{2}{c}{Condition} & \multicolumn{3}{c}{Distribution alignment} \\
   \cmidrule(rl){3-4} \cmidrule(rl){5-7} \cmidrule(rl){8-9} \cmidrule(rl){10-12}
 & & Acc(\%)$_\uparrow$ & DoA Error$_\downarrow$ & FD$_\downarrow$ & FAD$_\downarrow$ & KL$_\downarrow$ & Acc(\%)$_\uparrow$ & DoA Error$_\downarrow$ & FD$_\downarrow$ & FAD$_\downarrow$ & KL$_\downarrow$ \\ 
\cmidrule{2-12}
Ground-Truth & Human & 87.19 & 32.06$^\circ$ & - & - & - & 68.43 & 37.37 & -  & - & -\\
\cmidrule{2-12}

\multirow{1}{*}{Simulated} & Reference audio & 62.66 & 3.33$^\circ$ & 10.79 & 2.47 & 1.70 & 85.14 & 4.12$^\circ$ & 15.94 & 3.05 & 1.90 \\
\multirow{1}{*}{(Baseline)} & AudioLDM & 14.57 & \textbf{3.22}$^\circ$ & 32.62 & 4.67 & 2.80 & 37.29 & \textbf{3.07}$^\circ$ & 22.64 & 3.11 & 1.98 \\   
 & Tango & 35.58 & 3.92$^\circ$ & 23.03 & 8.04 & 2.05 & 52.00 & 3.39$^\circ$ & 11.80 & 5.37 & 1.85 \\  

\cmidrule{2-12}

\multirow{1}{*}{(Ours)} & Diff-SAGe  & \textbf{76.52} & 22.97$^\circ$ & \textbf{3.93} & \textbf{0.64} & \textbf{1.44} & \textbf{85.29} & 31.96$^\circ$ & \textbf{6.46} & \textbf{0.98} & \textbf{1.66} \\
 
\bottomrule
\end{tabular}
}
\vspace{-3mm}
\end{table*}

\begin{table}[t]
\centering
\caption{\normalfont{Ablation study on TAU-19.}}
\vspace{-3mm}
\label{tab:ablation}
\resizebox{\columnwidth}{!}{ 
\begin{tabular}{l|ccccc}
\toprule
 & \multicolumn{2}{c}{Condition} & \multicolumn{3}{c}{Distribution Alignment} \\
\cmidrule(rl){2-3} \cmidrule(rl){4-6}
  & Acc(\%)$_\uparrow$ & DoA Error$_\downarrow$ & FD$_\downarrow$ & FAD$_\downarrow$ & KL$_\downarrow$  \\ 
\midrule
 Ground-Truth & 87.19 & 32.06$^\circ$ & - & - & - \\
\midrule
 AudioLDM (sim) & 14.57 & 3.22$^\circ$ & 32.62 & 4.67 & 2.80  \\ 
\midrule

Diff-SAGe  & 76.52 & 22.97$^\circ$ & \textbf{3.93} & \textbf{0.64} & 1.44 \\

Diff-SAGe-B & \textbf{76.60} & 22.21$^\circ$ & 4.85 & 0.81 & \textbf{1.43} \\

Diff-SAGe-B (+ aug) & 70.71 & 16.96$^\circ$ & 7.42 & 1.20 & 1.48 \\

Diff-SAGe-B (sim) & 71.14 & \textbf{3.19}$^\circ$ & 10.82 & 2.22 & 1.74 \\
 
\bottomrule
\end{tabular}
}

\end{table}

\noindent
{\textbf{Datasets.}}
Sound event localization and detection (SELD) is a well-known task in machine listening. For this new task, we use SELD dataset labels and FOA recordings to construct our training data. We restrict to generating static, non-overlapping sound sources of 1 second duration.
Specifically, we utilize the TAU Spatial Sound Events 2019 (TAU-19)~\cite{adavanne2019multi} and TAU-NIGENS Spatial Sound Events 2020 (TAU-NIGENS-20)~\cite{politis2020dataset} datasets for our experiments. Both the datasets are generated by convolving real mono audio recordings~\cite{mesaros2017detection,trowitzsch2019nigens} with real RIRs. We use the mono recordings for our reference audio simulation baseline and for data augmentation. 
TAU-19 includes 11 office-related sound classes (e.g., door knock, keyboard typing) recorded in 5 distinct spaces. From the original split, we extract 1-second segments, resulting in 15,798/3,974 train/test data points. TAU-NIGENS-20, with greater spatial diversity, has 14 classes (e.g., baby crying, dog barking) recorded in 13 different locations. After removing overlapping and moving sources we obtain a train/test split of 14,078/700 samples. 

\noindent
{\textbf{Experimental Details.}}
We resample FOA audio to 16kHz and crop or pad each data point to 1-second duration. We create FOA spectrograms with $T=64$ and $F=128$. For modeling, we use SiT~\cite{peebles2023scalable} diffusion framework, utilizing default training parameters, including a constant learning rate of $1\times10^{-4}$ with Adam optimizer. Specifically, we use the SiT-B(ig) and SiT-L(arge) models, referred to as Diff-SAGe-B and Diff-SAGe, with parameter sizes of 132M and 462M, respectively. Our models are trained on 4$\times$A10 GPUs with a total batch size of 24 for Diff-SAGe-B and 16 for Diff-SAGe, over 500 epochs. During training, class and spatial location conditions are randomly dropped (independently) with a probability($p$) of 10\%. We use 250 sampling steps and apply classifier-free guidance with a CFG value($\zeta$) of 4.0.

\noindent
{\textbf{Evaluations.}} We measure model performance in terms of both objective metrics and subjective evaluation.

\subsubsection{\underline{Objective Metrics}}
To quantitatively evaluate the quality of generated spatial audio, we broadly focus on input conditioning and distribution alignment metrics. We generate (or simulate) data using the class and spatial location of the test data. For evaluating conditioning, we assess class accuracy and Direction-of-Arrival (DoA) error. A pre-trained mono-audio classifier is used to calculate class accuracy, while DoA is estimated by applying a decoding matrix at 900 uniformly distributed points on the unit sphere and evaluating the maximum of the steered power~\cite{Zotter2019}. We also evaluate widely-used mono audio generation metrics: Fréchet Distance (FD)~\cite{liu2023audioldm}, Fréchet Audio Distance (FAD)~\cite{kilgour2018fr}, and KL-Divergence (KL)~\cite{10112585}, by extracting the $a_w(t)$ channel. 
These distribution alignment metrics are computed on the test sets. 

\subsubsection{\underline{Subjective Evaluation}}
We conducted a user study with 12 participants, each evaluating 15 randomly selected samples from Diff-SAGe, Ground Truth, and AudioLDM (simulated), with 5 samples of the same class and spatial location from each. To evaluate DoA, FOA recordings were rendered to canonical 7.1.4 in a listening room, and the mean subjective DoA error was reported. Subjective DoA error is the angular distance between the input-conditioned spatial location and the human-estimated location. Following~\cite{zhang2024foleycrafter}, participants were shown pairs of mono model outputs and asked to choose the one with better class relevance and audio quality (or select both). The performance score is defined as $(S\times100/A)$, where $S$ is the number of times a model was selected, and $A$ is the total number of appearances.

\begin{table}[t]
\centering

\caption{\normalfont{Subjective tests.}}
\vspace{-3mm}
\label{tab:user}
\resizebox{0.92\columnwidth}{!}{%
\begin{tabular}{@{}l|cccc@{}}
\toprule
 & DoA Error$_\downarrow$  &  Class-Relevance$_\uparrow$  & Audio Quality$_\uparrow$   \\ \midrule
Ground-Truth & 29.74$^\circ$ & 96.16 & 91.66 \\
\midrule
AudioLDM  & 50.30$^\circ$ & 34.16 & 30.83  \\ 
Diff-SAGe & \textbf{37.93}$^\circ$ & \textbf{82.50} & \textbf{73.33} \\
\bottomrule
\end{tabular}%
}
\label{tab:user}
\end{table}
\section{Results}
\subsubsection{\underline{Comparison with Baselines}} Table~\ref{tab:main} compares the performance of Diff-SAGe with other baselines across two datasets. Our approach outperforms in most cases, demonstrating the effectiveness of our approach. Diff-SAGe can generate class-aligned and distinctive audios that outperform on accuracy. 
For both datasets, a high DoA error is shown in the ground truth, potentially due to human annotation errors. Though our model is trained with these labels, we can improve this DoA error by a large margin ($\sim$10° in TAU-19 and $\sim$6° in TAU-NIGENS-20). Still, there exists a large gap with respect to simulation-based baselines. We will explain this gap through our ablation studies. Furthermore, Diff-SAGe outperforms baselines in all distribution alignment metrics by a large margin, highlighting the realistic generation quality. 

\subsubsection{\underline{Ablation studies}}
\label{sec:ablation}
Table~\ref{tab:ablation} illustrate our ablation studies, in which we compare the effect of model size and analyze the DoA performance gap of our approach with the baselines. 
We find the benefit of a large model size, as Diff-SAGe achieves similar or better results than Diff-SAGe-B. Next, we try to improve the DoA by utilizing additional data by convolving reference mono audio data on RIRs generated by \texttt{SpatialScraper} ~\cite{roman2024spatial}. This results in 40,000 more samples with 10 more rooms. As shown, Diff-SAGe-B (+aug) reduces the DoA Error of Diff-SAGe-B by $\sim$6°. Next, we train Diff-SAGe-B on data simulated by extracting the $a_w(t)$ channel of our real training data, which is represented by Diff-SAGe-B (sim), and observe a large reduction in DoA error. This even surpasses the AudioLDM baseline, though significantly suffering in other metrics. Thus, we conclude that the DoA error of our approach was limited by the training data annotation errors.

\subsubsection{\underline{User Study}} We show the results of the subjective test in Tab.~\ref{tab:user}. The high DoA error across the board supports our claim that recognizing DoA precisely is a difficult task. Unlike the objective DoA error, subjective DoA evaluations prefer our model to simulated data. In addition, high class relevance and generation quality metrics highlight the efficacy of our approach.
\section{Conclusion}
In this work, we introduce a novel task of native spatial audio generation conditioned on both class category and spatial location. To address this task, we propose Diff-SAGe, an end-to-end flow-based diffusion-transformer model. Our approach incorporates a multi-conditional encoder and addresses the limitations of phase estimation commonly used in mono audio generation. Through extensive evaluation, we demonstrate that Diff-SAGe surpasses simulation-based baselines in both objective and subjective metrics.
Future research will focus on addressing current limitations, such as extending to longer audio durations, handling multiple simultaneous sources, and developing compact phase-preserving FOA representations.

\vfill \pagebreak
\bibliographystyle{IEEEtran}
\bibliography{IEEEabrv,refs}

\end{document}